\documentclass{icrc2009}

\usepackage{graphicx}   
\usepackage[caption=false]{caption}    
\usepackage[font=footnotesize]{subfig} 
\usepackage{fixltx2e}
\usepackage{url}

\newcommand{\shorttitle}[1]%
{\markboth{Proceedings of the 31\MakeLowercase{$^{st}$} ICRC, {\L}\'{o}d\'{z} 2009}{#1} }
\newcommand{\etal}{\MakeLowercase{\textit{et al. }}} 

\begin{document}
\title{High Quantum Efficiency Phototubes for Atmospheric Fluorescence Telescopes}

\author{\IEEEauthorblockN{Daniel Kruppke-Hansen\IEEEauthorrefmark{1} and
			  Karl-Heinz Kampert\IEEEauthorrefmark{1}}
                            \\
\IEEEauthorblockA{\IEEEauthorrefmark{1}Bergische Universit\"{a}t
			  Wuppertal, Gau\ss stra\ss e 20, 42097 Wuppertal, Germany}}

\shorttitle{D. Kruppke-Hansen \etal High QE PMTs}
\maketitle

\begin{abstract}
  The detection of atmospheric fluorescence light from extensive air
  showers has become a powerful tool for accurate measurements of the
  energy and mass of ultra-high energy cosmic ray particles. Employing
  large area imaging telescopes with mirror areas of 10\,m$^2$ or
  more, showers out to distances of 30\,km and more can be
  observed. Matrices of low-noise photomultipliers are used to detect
  the faint light of the air showers against the ambient night-sky
  background noise. The signal-to-noise ratio of such a system is
  found to be proportional to the square root of the mirror area times
  the quantum efficiency of the phototube. Thus, higher quantum
  efficiencies could potentially improve the quality of the
  measurement and/or lead to the construction of more compact
  telescopes.  In this paper, we shall discuss such improvements to be
  expected from high quantum efficiency phototubes that became
  available on the market only very recently. A series of simulations
  has been performed with data of different types of commercially
  available high quantum efficiency phototubes. The results suggest a
  higher aperture and thus increased statistics for such
  telescopes. Additionally, the quality of the reconstruction can be
  improved.
\end{abstract}

\begin{IEEEkeywords}
  PMTs, Quantum Efficiency, Fluorescence Telescopes
\end{IEEEkeywords}

\section{Introduction}
  The continuous development of photomultiplier tubes led, in the last
  few years, to a major progress for the photocathodes towards
  cathodes with higher quantum efficiencies. For instance today, such
  super or ultra bialkali photomultipliers can have efficiencies of
  over $40\,\%$ (compared to the ``old standard'' of $25\,\%-30\,\%$)
  and are now commercially available on the market.

  This change is particularly interesting for fluorescence telescopes,
  whose detection of atmospheric fluorescence light from extensive air
  showers has become a powerful tool for accurate measurements of the
  energy and the mass of ultra high energy cosmic ray
  particles. Telescopes with mirror areas of $10\,$m$^2$ and more allow
  the observation of showers out to distances of $30\,$km. For this
  task it is essential to have low-noise photomultipliers which are
  able to detect the faint fluorescence light from the shower. The
  signal-to-noise ration for such a system is found to be
  \begin{equation}
    S/N\sim\sqrt{A\cdot \varepsilon_{\mathrm{Q}}},
  \end{equation}
  where $A$ is the mirror area and $\varepsilon_{\mathrm{Q}}$ the quantum efficiency of the
  photomultipliers in the camera. Hence, higher quantum efficiencies
  could potentially improve the quality of the measurements and/or led
  to a compacter construction of telescopes.

  In this work we discuss important properties of such super-bialkali
  and ultra-bialkali photomultipliers and describe simulations of
  atmospheric fluorescence telescopes employing such photomultipliers.

  The major objectives for this simulations are the possible increase
  in the trigger efficiency, and thus the aperture, of the telescope
  and the possible better quality of the reconstruction.

  \begin{figure*}[!t]
    \centerline{\subfloat{\includegraphics[width=3.2in]{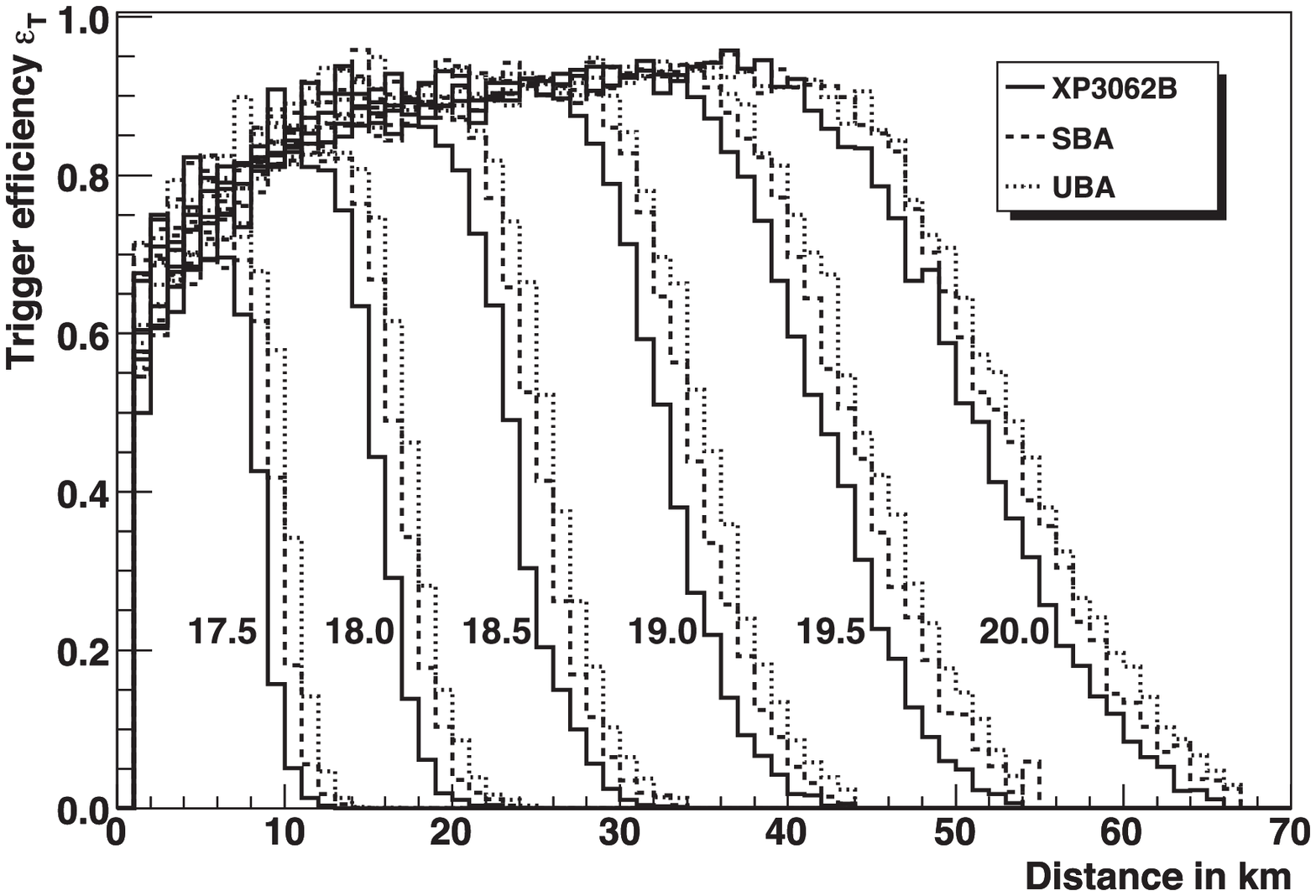} \label{triggerdistance}}
      \hfil
      \subfloat{\includegraphics[width=3.2in]{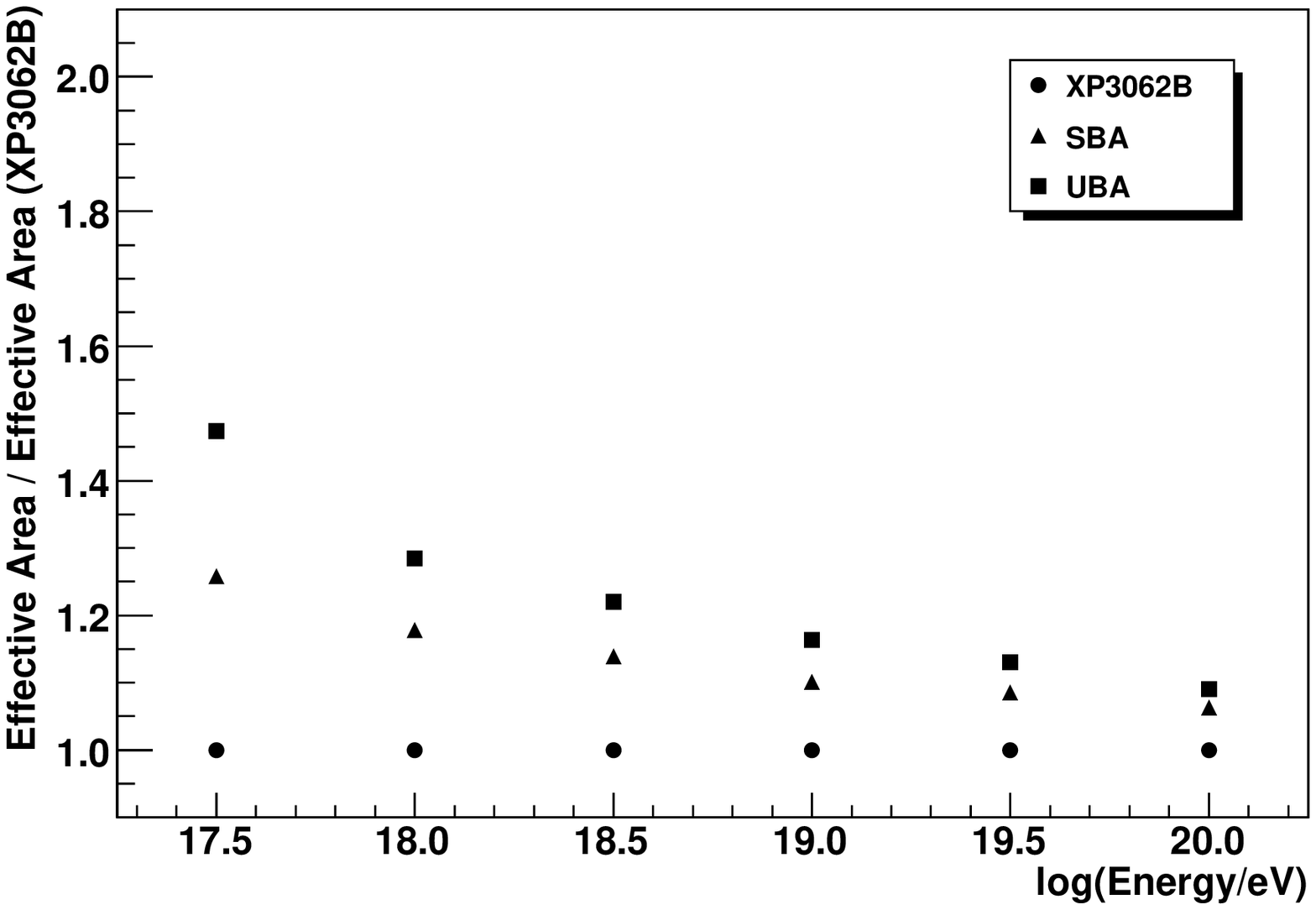} \label{effarea}}
    }
    \caption{The results of the trigger efficiency
    analysis. \emph{Left:} The dependence of the trigger efficiency on
    the distance from the telescope for different energies, indicated
    by $\log(\mathrm{Energy}/\mathrm{eV})$, and photomultipliers. The
    distance bin-size is $1\,$km. \emph{Right:} The dependence of the
    effective area on the energy and the different photomultipliers.}
    \label{triggerdistancefig}
  \end{figure*}

\section{Relevant Properties of High QE-PMTs}
  The task of the optical detector in fluorescence telescopes is to
  pick out the faint and fast signal of air showers from the ambient
  night-sky background noise which is of the order of 100 photons per
  $\mathrm{m}^2\,\mu\mathrm{s}\,\mathrm{deg}^2$ in the relevant
  wavelength range \cite{photonno}. Thus, besides the cathode and
  anode luminous sensitivities, also the anode dark current and after
  pulses are of particular importance together with the linearity and
  a high dynamic range. Very detailed information based on large
  quantities of XP3062B photomultipliers \cite{photonis} have been
  obtained from laboratory tests reported in \cite{massprod}. Such
  information is not yet available for super and ultra bialkali
  photomultipliers becoming available only very recently. Thus, a
  smaller number of super-bialkali photomultipliers has been procured
  and is now under study in the laboratory, while the ultra-bialkali
  photomultipliers presently appear to be too expensive for being
  considered a realistic option for large quantities as needed in a
  fluorescence detector camera.

  First tests aim for measuring the quantum efficiency as a function
  of wavelength in comparison to the XP3062B photomultipliers, the
  homogeneity of the photocathode response, the dark current,
  occurrence of after-pulses, and the dynamic range. When used in
  combination with MUG-6 filters in the aperture system, the absence
  of sensitivity to photons at wavelengths larger than
  $650\,\mathrm{nm}$ becomes another issue to be verified, as the
  filter becomes transparent again in this wavelength
  range. Preliminary studies show that the dark current of high
  quantum efficiency photomultipliers is increased compared to
  standard bialkali photomultipliers. This will be quantified and be
  reported together with other test-results at the conference.

\section{Simulations}
  The simulations involve three different steps, their details will be
  explained in this section: the shower simulation, the telescope
  simulation and the shower reconstruction.

  As sample for photomultipliers with different quantum efficiencies
  the data of:
  \begin{itemize}
    \item a Photonis XP3062B, which is the photomultiplier used for
    the Pierre Auger Observatory and the former HiRes experiment, with
    $\varepsilon_{\mathrm{Q}}\sim 28\,\%$ \cite{photonis},
    \item a Hamamatsu R9420-100 SBA (super-bialkali) with
    $\varepsilon_{\mathrm{Q}}\sim 35\,\%$ \cite{hamamatsu} and
    \item a Hamamatsu UBA (ultra-bialkali) photocathode with
    $\varepsilon_{\mathrm{Q}}\sim 43\,\%$ \cite{hamamatsu}
  \end{itemize}
  are used. For the second and third case only the data of the
  photocathodes, and not the, potentially, different gain of the
  multipliers, is important. This ensures an observation of effects
  only due to the quantum efficiency.

  \subsection{Shower simulation}
    For a fluorescence telescope the interesting part of an airshower
    is its longitudinal profile and the emitted fluorescence light,
    but not the whole particle content of the shower. For this purpose
    the simulation program CONEX \cite{conexone, conextwo} was
    chosen. It uses a combination of Monte Carlo simulations and
    solutions of the cascade equations for airshowers and is, thus, a
    very fast alternative to pure Monte Carlo simulation programs.

    The primary particles were chosen to be protons with an energy
    distribution uniformly in $\log(E)$ in steps of 0.5 from
    $10^{17.5}\,$eV to $10^{20}\,$eV. For each energy step 25\,000
    showers have been simulated with a uniform azimuth angle
    distribution and a zenith angle distribution according to
    $\mathrm{d}N/\mathrm{d}\cos\theta\sim\cos\theta$ between $0^\circ$ and $60^\circ$ to
    take into account the flat surface on which the showers will be
    distributed.

  \subsection{Telescope simulation}
    The actual simulations of the different photomultipliers is done
    in this step. For convenience, we use the example of the Pierre
    Auger Observatory, which has four fluorescence detector sites. To
    exclude effects coming from showers seen in more than one site,
    only one of these is simulated. For the simulations we used the so
    called Offline Framework \cite{offline}, which is a
    simulation and reconstruction tool developed and used for the
    Auger Observatory.

    The shower cores are distributed on a $5^\circ$ slice in front of
    the telescope. The maximum distance between shower core and
    telescope was chosen in a way that the trigger probability (for
    the standard photomultiplier) for a shower outside this radius is
    lower than $1\,\%$.

  \subsection{Shower reconstruction}
    The last step of the simulation chain is the reconstruction of the
    shower. This is again done by using the Offline framework, which
    then yields all the important quantities, such as the
    reconstructed profile, energy and the position of the shower
    maximum ($X_{\mathrm{max}}$).
    
\section{Results}

  \subsection{Trigger efficiency}
    The first expectation on photomultipliers with higher quantum
    efficiency is the possibility to observe fainter signals, which
    for airshowers correlates to the possibility of observing more
    distant showers. One observable to measure this effect is the
    trigger efficiency against the distance. As trigger efficiency we
    define the ratio $\varepsilon_\mathrm{T}=N_T/N_S$, the number of
    showers which trigger during the telescope simulation over the
    total number of simulated showers. This ratio is calculated for
    each energy-distance bin. The result is shown in
    figure~\ref{triggerdistance}. As expected, for all energies the
    trigger efficiency rises if the quantum efficiency of the
    photomultipliers becomes higher, or in other words, the telescope
    can see showers of $\sim 2\,\mathrm{km}$ farther away.

    For a clean atmosphere the expectation would be a factor of
    $r(\mathrm{new})/r(\mathrm{old})=\sqrt{\varepsilon_\mathrm{Q}(\mathrm{old})/\varepsilon_\mathrm{Q}(\mathrm{new})}$
    due to the $1/r^2$ intensity loss, valid for all energies. But Mie
    and Rayleigh scattering attenuate this effect for larger
    distances.

    To see the result in a more quantitative way it is useful to look
    at the effective areas defined as
    \begin{equation}
      A_{\mathrm{eff}}=\int\varepsilon(r)r\,\mathrm{d}r\,\mathrm{d}\varphi\sim\sum_r\varepsilon(r)r,
    \end{equation}
    where the sum runs over the different distance bins with a
    bin-size of $1\,$km. Figure~\ref{effarea} shows the relative
    effective areas compared to the one obtained for the standard
    photomultipliers. The increase for all energies is obvious and
    becomes larger for smaller energies because of a geometrical
    reason, i.\,e.\ the relative area growth of a slice is of the
    order $\mathcal{O}(\Delta r/r)$.

  \subsection{Number of reconstructed showers}
    \begin{figure}[!t]
      \centering
      \includegraphics[width=3.2in]{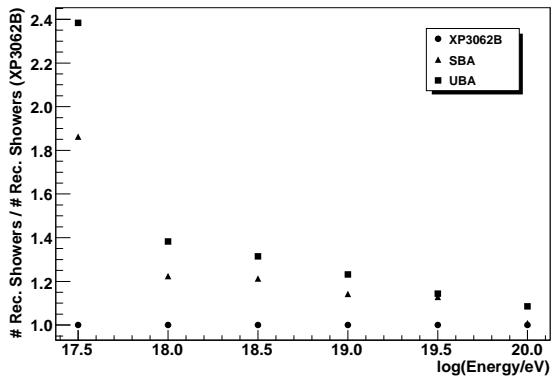}
      \caption{The relative number of reconstructed showers which pass
      the quality cuts, compared to the number for the standard
      photomultipliers.}
      \label{norec}
    \end{figure}

    Before having a look at the reconstructed showers, the shower
    sample has to be cleaned up by applying several quality cuts.  The
    main cuts are:
    \begin{itemize}
      \item the reconstructed shower maximum shall lie inside the
      observed energy deposit profile,
      \item the absolute uncertainty ($\Delta X_{\mathrm{max}}$) of
      the reconstructed shower maximum is smaller than
      $40\,\mathrm{g/cm}^2$,
      \item the relative uncertainty of the reconstructed energy
      ($\Delta E/E$) is smaller than $20\,\%$,
      \item the reduced $\chi^2$ of the Gaisser-Hillas fit is smaller
      than 5 and
      \item the estimated amount of Cherenkov light at the aperture
      should be smaller than $50\,\%$.
    \end{itemize}

    Since the effective area increases for higher quantum
    efficiencies, we expect to have more ``good'' reconstructed
    showers in theses cases. The result in figure~\ref{norec} shows a
    significant increase of the relative number of successfully
    reconstructed showers. For lower energies this effect becomes
    larger, which is correlated to the behaviour of the effective area
    described in the last section.

  \subsection{Gaisser-Hillas profile}
    \begin{figure}[!t]
      \centering
      \includegraphics[width=3.2in]{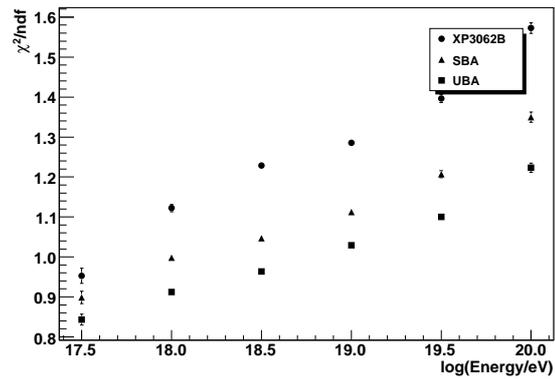}
      \caption{The $\chi^2/ndf$ distributions of the Gaisser-Hillas
	fits. Shown are the means of the distributions and their
	corresponding errors.}
      \label{gh}
    \end{figure}

    A good indicator for the quality of the reconstruction is the
    goodness of the Gaisser-Hillas fit, which gives the shower
    profile. The goodness is described by the $\chi^2/ndf$ of the
    fit. Figure~\ref{gh} shows the means of the different $\chi^2/ndf$
    distributions as function of the energy. As can be seen the,
    higher quantum efficiencies yield a significant improvement of the
    goodness of the fits and thus of the quality of the
    reconstruction. This effect is as expected because higher quantum
    efficiencies should yield more signals in the camera and thus
    facilitate the reconstruction.

  \subsection{Energy and $X_{\mathrm{max}}$ resolution}
  \begin{figure*}[!t]
    \centerline{\subfloat{\includegraphics[width=3.2in]{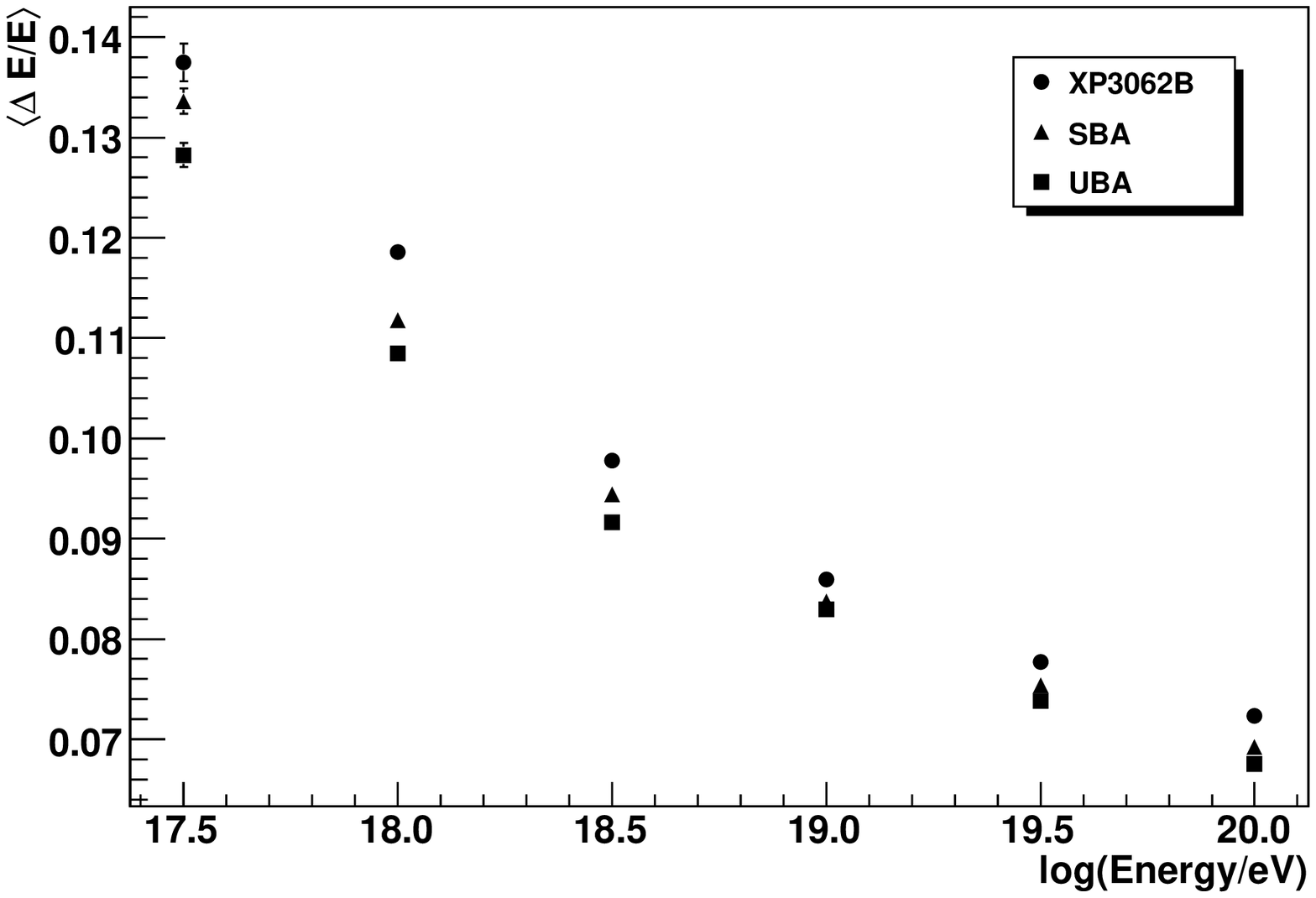} \label{energyerr}}
      \hfil
      \subfloat{\includegraphics[width=3.2in]{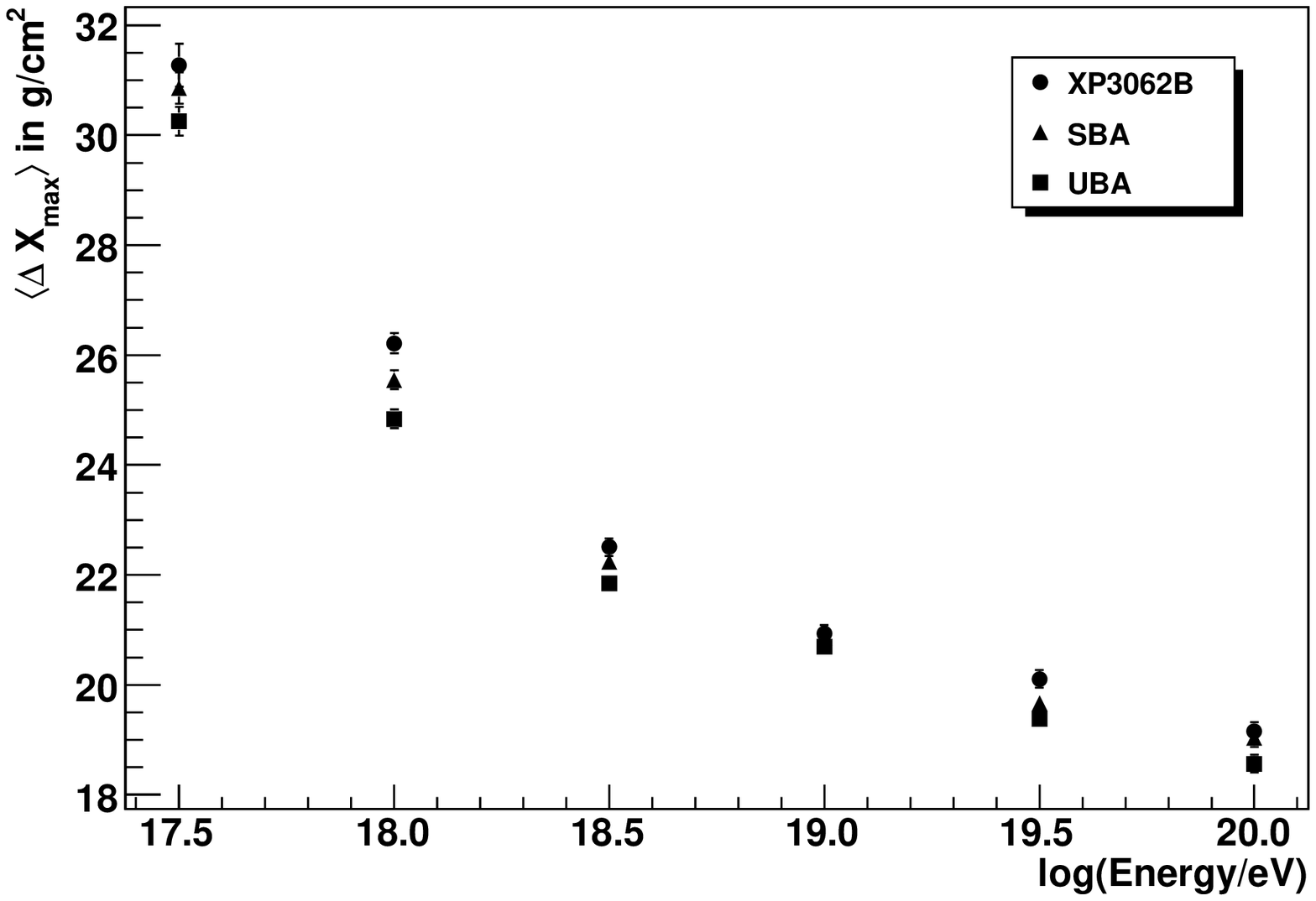} \label{xmaxerr}}
    }
    \caption{\emph{Left:} The energy resolutions for the different quantum
	efficiencies. \emph{Right:} The $X_{\mathrm{max}}$ resolutions for the different
	quantum efficiencies. In both cases the means and their
	corresponding errors of the different distributions are shown.}
    \label{triggerdistancefigx}
  \end{figure*}

    Since the Gaisser-Hillas fit becomes significantly better for
    photomultipliers with higher quantum efficiencies, we can assume
    that also the resolution of the quantities derived from the
    profile will become better. Namely, these are the energy and the
    position of the shower maximum ($X_{\mathrm{max}}$). Their mean
    residuals are shown in figure~\ref{energyerr} and
    figure~\ref{xmaxerr}, respectively. As expected they tend to lower
    values and thus to more precise energy and $X_{\mathrm{max}}$
    reconstructions.

\section{Discussion and Outlook}
  The results of the simulations done in this work clearly show that
  photomultipliers with higher quantum efficiencies can significantly
  improve the quality of airshower measurements. Furthermore, they
  allow the observation of more distant showers or of showers with
  lower energies at given distance. Both of this increases the aperture
  of the telescopes.

  The simulation of these new photomultipliers with high quantum
  efficiencies is, of course, only the first step towards a real usage
  in an experiment. As a next step samples of these photomultipliers
  are currently tested to determine their properties under conditions
  typically for a fluorescence telescope. A final step might then be
  the equipment of a whole telescope with these new photomultipliers.

\section*{Acknowledgement}
  We gratefully acknowledge financial support of the German Federal Ministry of Education
  and Research (BMBF grant 05 A0PX1).

  In addition, we would like to thank the Pierre Auger Collaboration for the usage
  of the Offline Framework.

\end{document}